\documentclass[12pt,epsf]{article}
\usepackage{amsmath}
\usepackage{amssymb}
\begin{document}
%%%%%%%%%%%%%%%%%  Defs. %%%%%%%%%%%%%%%%%%%%%%%%%%%%%%
\def\a{\alpha}
\def\b{\beta}
\def\c{\varepsilon}
\def\d{\delta}
\def\e{\epsilon}
\def\f{\phi}
\def\g{\gamma}
\def\h{\theta}
\def\k{\kappa}
\def\l{\lambda}
\def\m{\mu}
\def\n{\nu}
\def\p{\psi}
\def\q{\partial}
\def\r{\rho}
\def\s{\sigma}
\def\t{\tau}
\def\u{\upsilon}
\def\v{\varphi}
\def\w{\omega}
\def\x{\xi}
\def\y{\eta}
\def\z{\zeta}
\def\D{{\mit \Delta}}
\def\G{\Gamma}
\def\H{\Theta}
\def\L{\Lambda}
\def\F{\Phi}
\def\P{\Psi}

\def\S{\Sigma}

\def\o{\over}
\def\beq{\begin{eqnarray}}
\def\eeq{\end{eqnarray}}
\newcommand{\gsim}{ \mathop{}_{\textstyle \sim}^{\textstyle >} }
\newcommand{\lsim}{ \mathop{}_{\textstyle \sim}^{\textstyle <} }
\newcommand{\vev}[1]{ \left\langle {#1} \right\rangle }
\newcommand{\bra}[1]{ \langle {#1} | }
\newcommand{\ket}[1]{ | {#1} \rangle }
\newcommand{\EV}{ {\rm eV} }
\newcommand{\KEV}{ {\rm keV} }
\newcommand{\MEV}{ {\rm MeV} }
\newcommand{\GEV}{ {\rm GeV} }
\newcommand{\TEV}{ {\rm TeV} }
\def\slash#1{\ooalign{\hfil/\hfil\crcr$#1$}}
\def\diag{\mathop{\rm diag}\nolimits}
\def\Spin{\mathop{\rm Spin}}
\def\SO{\mathop{\rm SO}}
\def\O{\mathop{\rm O}}
\def\SU{\mathop{\rm SU}}
\def\U{\mathop{\rm U}}
\def\Sp{\mathop{\rm Sp}}
\def\SL{\mathop{\rm SL}}
\def\tr{\mathop{\rm tr}}

%%%%%%%%%%%%%%%%%%%%%%%%%%%%%%%%%%%%%%%%%%%%%%%%%%%
%%%%%%%%%%%%%%%%%%%%%%%%%%%%%%%%%%%%%%%%%%%%%%%%%%%
\baselineskip 0.7cm

\begin{titlepage}
\begin{flushright}
UCB-PTH-07/23
\end{flushright}

\begin{center}

\vskip 2.35cm

{\large \bf Stable SUSY Breaking Model with $O(10)$ eV Gravitino}
{\large \bf from  Combined D-term Gauge Mediation and $U(1)'$ Mediation}
\vskip 1.2cm
\end{center}

\begin{center}
{\large \bf }
\vskip 1.2cm

Yu Nakayama${}^{1}$

\vskip 0.8cm

${}^1${\it Berkeley Center for Theoretical Physics and Department of Physics, 
\\
University of California, Berkeley, California 94720-7300}

\vskip 1.4 cm

\abstract{We show a calculable example of stable supersymmetry (SUSY) breaking models with $O(10)$ eV gravitino mass based on the combination of D-term gauge mediation and  $U(1)'$ mediation. A potential problem of the negative mass squared for the SUSY standard model (SSM) sfermions in the D-term gauge mediation is solved by the contribution from the $U(1)'$ mediation. On the other hand, the splitting between the SSM gauginos and sfermions in the $U(1)'$ mediation is circumvented by the contributions from the D-term gauge mediation. Since the $U(1)'$ mediation does not introduce any new SUSY vacua, we achieve a completely stable model under thermal effects. Our model, therefore, has no cosmological difficulty.}
\end{center}
\end{titlepage} 

\setcounter{page}{2}

%%%%%%%%%%%%%%%%%%%%%%%%%%%%%%%%%%%%%%%%%%%%%%%%%%%%%%%%%%%%%%%%%%%%%%%%%%%%%%%%%%%%%%%%%%%%%%%%%%%%%%
\section{Introduction}
One definite prediction of the low energy supersymmetry (SUSY) breaking is the existence of the gravitino. Known as ``gravitino problem", however, the gravitino mass shows a quite severe constraint from the cosmology unless we introduce a sophisticated modification of the cosmological scenario \cite{Pagels:1981ke}\cite{K1}\cite{K2}\cite{Moroi:1993mb}\cite{K3}\cite{de Gouvea:1997tn}\cite{K4}. For example, if the gravitino is unstable with long life-time, we have to face its potential disturbance on the big bang nucleosysnthesis through its decay. If the gravitino is stable, on the other hand, we have to face the over-closure of the universe from its thermal production. A study of the dark matter \cite{Viel:2005qj} suggests that when the gravitino contributes to the total matter density of the universe as a warm dark matter with the existence of an independently assumed cold dark matter, the gravitino mass $m_{3/2}$ is bounded above as $m_{3/2} \lsim 16$ eV.

The search of the SUSY breaking mediation scenario that realizes $m_{3/2} \lsim 16 $ eV is, therefore, of cosmological significance (see e.g. \cite{Izawa:1997gs}\cite{Izawa:1999vc}\cite{Izawa:2005yf}\cite{Ibe:2007wp}\cite{Ibe:2007ab}\cite{INY} for some attempts in this direction). In our recent paper \cite{Nakayama:2007cf}, we proposed a strongly coupled D-term gauge mediation, where the very light gravitino mass  of order $10$ eV is naturally realized. A potential problem of the scenario is that we could not compute the signature of the mass squared for the SUSY standard model (SSM) sfermions, and in the perturbative regime, the D-term gauge mediation predicts tachyonic mass squared \cite{Poppitz:1996xw}\cite{Nakayama:2007cf}. While there is no theoretical evidence for or against negative mass squared for the SSM sfermions in the strongly coupled D-term gauge mediation scenario, it would be certainly better to show a recipe to  remedy this potential problem. In this paper, we provide this recipe by introducing an extra ingredient.

The extra ingredient we add is $U(1)'$ mediation \cite{Langacker:2007ac}. We gauge a global $U(1)'$ symmetry of the SSM sector, and the $U(1)'$ gaugino acquires soft mass due to the $F$-term vacuum expectation value (VEV) in the hidden sector. In this scenario, the SSM sfermions obtain soft mass squared at the one-loop order while the SSM gauginos obtain soft mass at the two-loop order. Thus, the minimal $U(1)'$ mediation has its own problem of the splitting SUSY spectrum \cite{Langacker:2007ac}. 

Our strategy is to combine the contributions of the D-term gauge mediation and the contributions of the $U(1)'$ mediation. Each contributions solve the weakness of the other scenario complementarily. In addition, since both the D-term gauge mediation and the $U(1)'$ mediation do not introduce any unwanted SUSY preserving vacua in the global SUSY limit, our model is completely stable under the thermal effects. In this way, we realize a new class of SUSY breaking mediation scenarios with $O(10)$ eV gravitino mass, which is free from any cosmological difficulties.

Our construction is based on the field theory analysis, but it also has a natural embedding in the string theory. As a final appetizer, we would like to taste a flavor of such a possibility here. Indeed, the string theory has all the ingredients needed for the D-term gauge mediation and the $U(1)'$ mediation. The D-term SUSY breaking and its mediation can be easily embedded in the string theory as discussed in \cite{Nakayama:2007cf}, and an important fact is that any D-term SUSY breaking requires the F-term SUSY breaking in the string theory, which will be crucial for the $U(1)'$ mediation. The ingredient for the $U(1)'$ mediation can also be found in the string theory \cite{Blumenhagen:2005mu}\cite{Blumenhagen:2006ci}\cite{Verlinde:2007qk} because the global symmetry will always be gauged in the string theory. We will briefly sketch how stringy setup combines these two mediation scenarios. The detailed construction including the moduli stabilization will be investigated elsewhere.

The menu of the paper is as follows. In section 2, we present our scheme to combine the D-term gauge mediation and the $U(1)'$ mediation in a completely calculable regime. Although we cannot achieve $O(10)$ eV gravitino mass in this perturbative regime, we see how the difficulty of the each model is solved by the contributions of the other mediation. In section 3, we show a new class of the mediation scenario  with $O(10)$ eV gravitino mass by combining the non-perturbative D-term gauge mediation and the $U(1)'$ mediation. The thermal stability of the model is also investigated. In section 4, we give some discussions of our model, and in particular, we sketch the realization of our scenario in the string theory. In the appendix, we give a possible origin of the $U(1)'$ gaugino mass from the $U(1)'$ gauge mediation.

\section{Perturbative model with calculability}
We first present a new class of SUSY breaking mediation scenario based on the combination of the D-term gauge mediation \cite{Nakayama:2007cf} and the $U(1)'$ mediation \cite{Langacker:2007ac} in a perturbative regime. The individual phenomenological difficulties of the D-term gauge mediation and the $U(1)'$ mediation are solved complementarily. Furthermore, the model is completely calculable once we specify a dynamical SUSY breaking mechanism in the hidden sector.

Our assumption in the hidden sector is quite simple. We assume that the SUSY is broken both by F-term VEV $\langle F_S \rangle$ for a certain chiral superfield $S = \langle S\rangle  + \theta^2 \langle F_S \rangle $ and by $U(1)_D$ D-term VEV $\langle D \rangle $ (e.g. by constant or field dependent effective Fayet-Iliopoulos (FI) D-term). The coexistence of the F-term SUSY breaking and the D-term SUSY breaking is crucial in the following, and we just point out that in the supergravity, the existence of the F-term SUSY breaking is necessary for the existence of the D-term SUSY breaking.\footnote{This is due to the supergravity relation: $\sum_i \delta \phi_i^\dagger  \frac{F^i}{W} = D$, where $\delta \phi_i$ is the variation under the $U(1)_D$ gauge symmetry \cite{Choi:2005ge}.} The SUSY breaking is presumably induced by a strong dynamics of the hidden sector, and accordingly we  naturally set $\langle F_S \rangle \sim \langle D \rangle \sim \Lambda_S^2$, where $\Lambda_S$ is a typical dynamical scale of the hidden sector.\footnote{A concrete field theory realization of the dynamical generation of the D-term together with F-term SUSY breaking can be found in \cite{NYY2}. String theory idea to fix the D-term as well as the F-term has been studied in \cite{Nakayama:2007du}\cite{NP}.} 

The SUSY breaking effects in part are mediated as D-term gauge mediation by the messenger $f$ and $\bar{f}$ which are charged under the $U(1)_D$ and the standard model (GUT) group. For definiteness, we assume that $f$ transforms as $\mathbf{5}$ under the GUT group with $U(1)_D$ charge $+1$ and $\bar{f}$ transforms as $\mathbf{\bar{5}}$ under the GUT group with $U(1)_D$ charge $-1$. The messengers have supersymmetric mass $M$ from a superpotential $W_{\mathrm{mess}} = M f \bar{f}$. The leading two-loop contribution to the soft mass squared for SSM sfermions $m_i$ was computed to be \cite{Poppitz:1996xw}\cite{Nakayama:2007cf}
\begin{align}
m_i^2  &= - \frac{7}{9} \frac{g_{GUT}^4}{(16\pi^2)^2} \frac{D^4}{M^6} \cr
 &= - \frac{7}{9} \frac{g_{GUT}^4}{(16\pi^2)^2} \frac{\Lambda_S^8}{M^6} \ . \label{dgsfm}
\end{align}
 As discussed in the introduction, this contribution alone leads to a difficulty of negative sfermion mass squared.
Furthermore, there is no direct generation of soft mass for SSM gauginos from the (minimal) D-term gauge mediation scenario considered here.

In our scenario, we also introduce the $U(1)'$ mediation recently studied in \cite{Langacker:2007ac}. We gauge a $U(1)$ global symmetry of the SSM, which will be denoted by $U(1)'$. It can be non-anomalous (i.e. $U(1)' \equiv U(1)_{B-L}$) as well as anomalous. The $U(1)'$ will be broken at the scale $M_{Z'}$ by the vacuum expectation value of $U(1)'$ charged fields or by the four-dimensional Green-Schwarz mechanism.\footnote{These extra contributions will cancel the anomaly of $U(1)'$ if any.} We couple the SUSY breaking singlet chiral superfield $S$ to the $U(1)'$ gauge kinetic term as
\begin{eqnarray}
\int d^2\theta \frac{S}{m} W^\alpha W_{\alpha} + h.c. \ , \label{mass}
\end{eqnarray}
which will induce the non-supersymmetric $U(1)'$ gaugino mass $\frac{{F}_S}{m} \sim \frac{\Lambda_S^2}{m}$. We have assumed here and continue to assume hereafter $M_{Z'} \ll \frac{\Lambda_S^2}{m}$ so that the supersymmetric mass for the $U(1)'$ vector multiplet may be neglected. This assumption also suppresses the contribution of the $U(1)'$ D-term to the soft mass squared for SSM sfermions, which we will neglect in the following. We also neglect all the other non-renormalizable operators possibly suppressed only by $(1/m)^2$ (see the appendix for a further discussion on this point).

Due to the mass difference in the $U(1)'$ vector multiplet, SSM sfermions acquire one-loop mass squared
 as \cite{Langacker:2007ac}
\begin{align}
m_i^2 &\sim \frac{g_{Z'}^2}{16\pi^2} \left(\frac{F_S}{m}\right)^2 Q_i^2 \log\left(\frac{\Lambda_S m}{{F_S}}\right) \cr
 &= \frac{g_{Z'}^2}{16\pi^2} \left(\frac{\Lambda_S^2}{m}\right)^2 Q_i^2 \log\left(\frac{m}{\Lambda_S}\right) \ , \label{xsfm}
\end{align}
where $g_{Z'}$ is the coupling constant for the $U(1)'$ and $Q_i$ is the $U(1)'$ charge of the SSM field that will be assumed to be $O(1)$. The flavor changing neutral current problem is avoided when $U(1)'$ is flavor blind like $U(1)_{B-L}$. Furthermore, the SSM gaugino acquires two-loop mass as \cite{Langacker:2007ac}
\begin{align}
m_{1/2} &\sim \frac{g_{Z'}^2 g_{GUT}^2}{(16\pi^2)^2} \left(\frac{F_S}{m}\right) \log\left(\frac{\Lambda_Sm}{F_S} \right) \cr
 &= \frac{g_{Z'}^2 g_{GUT}^2}{(16\pi^2)^2} \left(\frac{\Lambda_S^2}{m}\right) \log\left(\frac{m}{\Lambda_S} \right) \ .
\end{align}
As discussed in the introduction, the $U(1)'$ mediation itself leads to a splitting SUSY spectrum: $m_{1/2} \sim \frac{g_{Z'}g_{GUT}^2}{(16\pi^2)^{3/2}} m_{i}$ and we have to abandon the SUSY solution of the hierarchy problem, at least partially.

Our proposal is that we can remedy these individual problems complementarily by combining the D-term gauge mediation and the $U(1)'$ mediation. The combined contributions from \eqref{dgsfm} and \eqref{xsfm} to the mass squared for SSM sfermions  are
\begin{align}
m_i^2 = - \frac{7}{9} \frac{g_{GUT}^4}{(16\pi^2)^2} \frac{\Lambda_S^8}{M^6} + \frac{g_{Z'}^2}{16\pi^2} \left(\frac{\Lambda_S^2}{m}\right)^2 Q_i^2 \log\left(\frac{m}{\Lambda_S}\right) \ ,\label{combsfm}
\end{align}
and the combined mass for SSM gaugino is
\begin{align}
m_{1/2} = 0 + \frac{g_{Z'}^2 g_{GUT}^2}{(16\pi^2)^2} \left(\frac{\Lambda_S^2}{m}\right) \log\left(\frac{m}{\Lambda_S} \right) \ .
\end{align}
Now, because of the negative mass squared contribution from the D-term gauge mediation, one can balance the first term and the second term in \eqref{combsfm} so that $m_i \sim m_{1/2}$.\footnote{Actually, this is quite a technical tuning because of the different renormalization group running of the gauge coupling constant of the SSM gauge group below the GUT scale. We have to tune $Q_i$ judiciously, depending on the SSM gauge group quantum number, so that the cancellation occurs for each SSM sfermion. Such unnatural tuning will be avoided in the next section, where the non-perturbative D-term gauge mediation is considered.}

While precise cancellation requires a tuning of $Q_i$, the rough cancellan yields a condition
\begin{eqnarray}
\frac{g_{Z'}^2}{16\pi^2} \frac{\Lambda_S^4}{m^2} \sim \frac{g_{GUT}^4}{(16\pi^2)}\frac{\Lambda^8_S}{M^6} \ ,
\end{eqnarray}
or
\begin{eqnarray}
\frac{\Lambda_S^2} {m} \sim 10^{-1} \frac{\Lambda^4_S}{M^3} \ , 
\end{eqnarray}
where we have assumed $g_{Z'} \sim g_{GUT}$.
If we demand that the SSM gaugino mass is $m_{1/2}\sim 10^3$ GeV, we have to set $\frac{\Lambda_S^2}{m} \sim 10^7$ GeV. It is then possible to obtain $m_i \sim 10^3$ GeV by the cancellation discussed above.

Let us investigate the gravitino mass $m_{3/2} \sim \frac{\Lambda_S^2}{M_{pl}} \sim 10^7 \left(\frac{m}{M_{pl}}\right)$ GeV of the model.\footnote{We set $M_{pl} = 2.4 \times 10^{18}$ GeV as  (reduced) Planck mass.} The largest $m$ would be $m \sim M_{pl}$, and it would lead to $M_{3/2} \sim 10^7$ GeV.\footnote{In such a scenario, however, the effects of the gravity mediation, in general, dominate over the effects considered here, so a sequestering mechanism to suppress the gravity mediation (and even the anomaly mediation) is needed.} A possible range of $m$ is bounded by $m > \Lambda_S$, and the smallest possible choice $m \sim \Lambda_S \sim 10^7$ GeV leads to the smallest gravitino mass for the current scenario: $m_{3/2} \sim 100$ keV. In this typical parameter region, gravitino is the lightest supersymmetric particle (LSP) and the constraint from the cosmology is quite severe \cite{Pagels:1981ke}\cite{Viel:2005qj}. The situation will be improved in the next section where we study the non-perturbative D-term gauge mediation.

Before going into the construction of the combined non-perturbative D-term gauge mediation and $U(1)'$ mediation, we have several comments on the perturbative model studied in this section. Actually, there are many different ways to solve the negative SSM sfermion mass squared problem once we allow F-term SUSY breaking in the hidden sector. One of the easiest possibilities would be to incorporate the F-term gauge mediation by adding an interaction $\int d^2\theta S f \bar{f}$. Indeed, such a term alone will serve as an excellent model of the SUSY breaking mediation, which is well-known as F-term gauge mediation (see e.g. \cite{Giudice:1998bp} for a review). The motivation to forbid such interaction here is  to avoid typical emergence of the SUSY preserving vacua after adding $\int d^2 \theta S f \bar{f}$. In the typical parameter region discussed here, such vacua are far away from the metastable SUSY breaking vacuum and they are quite irrelevant. However, when we would like to obtain a very light gravitino mass $\sim O(10)$ eV as we will achieve in the next section, it will become a problem, especially in the early universe with high temperature. We will avoid the new emergence of the SUSY preserving vacua by excluding the F-term gauge mediation. This will be important in the next section.

In this discussion, the origin of the interaction \eqref{mass} is crucial.\footnote{The author would like to thank M.~Ibe for the discussion.} If it were originated from the conventional $U(1)'$ F-term gauge mediation, it would give rise to the same problem of the thermal destabilization. One interesting observation is that we can  introduce an axionic shift symmetry of the chiral superfield $S$ in the imaginary direction: $S \to S + ix$ with a real number $x$. The effective SUSY breaking superpotential $W = \Lambda^2_S S + \frac{S}{m}W^\alpha W_\alpha$ satisfies this symmetry after integrating over the superspace. Note that we cannot replace the $U(1)'$ gauge field with the $SU(5)$ GUT gauge field here. The symmetry could remove the existence of the SUSY preserving vacua even in the ultra-violet theory. See appendix for a particular construction. In the string theory, such axionic coupling is indeed natural.

\section{Non-perturbative model with $O(10)$ eV gravitino}
Having explained our idea to combine the D-term gauge mediation and the $U(1)'$ mediation in the completely calculable perturbative example, we would like to propose a non-perturbative model with $O(10)$ eV gravitino mass. The model does not have SUSY preserving vacua, and hence does not possess any instability from the thermal effects in the early universe unlike the model in \cite{Ibe:2007ab}. Due to the non-perturbative nature of the strongly coupled D-term gauge mediation, the precise soft mass parameters will not be computed beyond the order estimation, but we emphasize that a possible negative SSM sfermion masses is guarantied to be avoided in our model unlike the ones discussed in \cite{Ibe:2007wp}\cite{Izawa:2005yf}\cite{Ibe:2007wp}\cite{Nakayama:2007cf}\cite{INY}.

The strongly coupled D-term gauge mediation proposed in \cite{Nakayama:2007cf} requires an additional strongly coupled $SU(N)$ gauge symmetry in the hidden sector in addition to the $U(1)_D$ D-term introduced in section 2. The messenger $f$ and $\bar{f}$ now transforms as $N$ (and $\bar{N}$) under the $SU(N)$ gauge symmetry. The strong dynamics of the $SU(N)$ gauge group leads to two important consequences: the one is that because of the gaugino condensation of the $SU(N)$ gauge group, the $R$ symmetry is broken and it is now feasible for the SSM gaugino to acquire soft mass within the D-term gauge mediation; the other is that the strong dynamics invalidates the perturbative computation of the SSM sfermion mass squared, which leads to the possibility that the negative mass squared for the SSM sfermions might be avoided. The latter ``hope" is actually unimportant in our model because the $U(1)'$ mediation automatically yields positive mass squared. For this purpose, the strongly coupled scale $\Lambda_N$ for the $SU(N)$ gauge group should be close to the messenger scale $M$: $32\pi^2 \Lambda_N^3 \sim M^3$.\footnote{This might be naturally realized by the strongly coupled conformal dynamics above the messenger scale: see \cite{Ibe:2007wp}.}

We now combine the $U(1)'$ mediation to the non-perturbative D-term gauge mediation. Both the non-perturbative D-term gauge mediation and the $U(1)'$ mediation gives rise to the soft mass squared for the SSM sfermions:
\begin{align}
m_i^2 &= \kappa \left(\frac{g_{GUT}^2}{16\pi^2}\right)^2 \frac{D^4}{M^6} +\frac{g_{Z'}^2}{16\pi^2} \left(\frac{F_S}{m}\right)^2 Q_i^2 \log\left(\frac{\Lambda_S m}{{F_S}}\right) \cr
 &=  \kappa_1 \left(\frac{g_{GUT}^2}{16\pi^2}\right)^2 \frac{\Lambda_S^8}{M^6} + \frac{g_{Z'}^2}{16\pi^2} \left(\frac{\Lambda_S^2}{m}\right)^2 Q_i^2 \log\left(\frac{m}{\Lambda_S}\right) \ ,
\end{align}
where $\kappa_1$ is an uncalculable $O(1)$ constant which could be either negative or positive.

Similarly the SSM gaugino mass acquires contributions both from the non-perturbative D-term gauge mediation and the $U(1)'$ mediation:
\begin{align}
m_{1/2} &= \kappa_2 \frac{g_{GUT}^2}{16\pi^2}\frac{D^4}{M^7} + \frac{g_{Z'}^2g_{GUT}^2}{(16\pi^2)^2} \frac{F_S}{m}  \cr
&=\kappa_2 \frac{g_{GUT}^2}{16\pi^2}\frac{\Lambda_S^8}{M^7} + \frac{g_{Z'}^2g_{GUT}^2}{(16\pi^2)^2} \frac{\Lambda_S^2}{m} \ ,
\end{align}
where $\kappa_2$ is another uncalculable $O(1)$ constant.

In order to avoid a splitting SUSY spectrum without resorting fine-tuning of parameters in the hidden sector, as we did in the last section, we should take $M\sim \Lambda_S$. The negative mass squared problem for SSM sfermions due to possible negative signature of $\kappa_1$ can be compensated by the positive mass squared contributions from the $U(1)'$ mediation by setting $\Lambda_S \sim 10^{-1} m$. The fine-tuning is less necessary than in the model discussed in section 2 because the SSM gaugino mass from the non-perturbative D-term gauge mediation is already comparable to the absolute value of the SSM sfermion mass from the D-term gauge mediation. 

This model allows a wide range of possible SSM soft parameters, but we focus on  the possibility of obtaining very light gravitino. If we demand that the SSM gauginos and sfermions are around $10^3$ GeV, $\Lambda_S$ should be around $10^5$ GeV. Then, the gravitino mass $m_{3/2} \sim \frac{\Lambda_S^2}{M_{pl}}$ is indeed as small as $O(10)$ eV. 

The SUSY breaking vacuum considered here is completely stable once we assume that the dynamical SUSY breaking sector is stable.\footnote{Explicit field theory realization of the stable dynamical SUSY breaking  based on the IYIT model with both F-term and D-term VEV can be found in \cite{NYY2}.} Unlike many F-term gauge mediation scenarios, the introduction of the messenger does not give rise to the emergence of SUSY preserving vacua. This is not a problem for high-scale gauge mediation because the potential barrier is high enough to prevent the thermal destabilization. However, in the very light gravitino scenario we are pursuing, the potential barrier is typically too low to be compatible with high scale inflation and the subsequent reheating. The absence of the SUSY preserving vacua is a sufficient condition to avoid the thermal destabilization.

\section{Discussion}
In this paper, we have proposed a new class of SUSY breaking mediation scenario based on the combination of the D-term gauge mediation and the $U(1)'$ mediation. The contributions of the each sector remedy the original problem of the individual scenario: negative mass squared for the SSM sfermion and splitting SUSY spectrum respectively. One feature of the model is that we can realize very light gravitino mass of order $10$ eV with no thermal instability. Almost all known models with this feature are based on the strong dynamics and the positive mass squared for the SSM sfermion was not guaranteed. The present model provides a concrete example with assured positive SSM sfermion mass squared. We would also like to mention that for a particular choice of the gauge group in the hidden sector, namely $N=5$, we have a candidate for a cold dark matter as composite baryon messengers \cite{Hamaguchi:2007rb}, which solves yet another cosmological difficulty of the usual low scale gravitino mass scenario. In this way, our model is cosmologically quite appealing.

Our model predicts massive $U(1)'$ vector multiplets (and possible exotics) at the scale around (or below) $10^{4}$ GeV. The mixing between the $U(1)'$ gauge boson and $Z$ leads to a lower bound on $M_{Z'}$ as $M_{Z'} \gsim (0.5\sim 1) $ TeV (see e.g. \cite{Everett:2000hb}). A phenomenology of these $U(1)'$ multiplets would be interesting.

Our model requires some tuning of the dimensional parameters such as a mass for the messenger and the dynamical scale of the hidden sector.
We note, however, that all the mass scales of the model: $\Lambda_S$, $M$, $\Lambda_N$, and $m$ are around $10^5 \sim 10^6$ GeV in the very light gravitino mass regime studied in section 3. Thus, there is a possibility that all these mass scales have a common origin, perhaps due to the hidden conformal dynamics or common source of the strong dynamics. It would be very interesting to construct explicit realization of this possibility.

Finally, we would like to discuss a possible stringy realization of the model. In \cite{Nakayama:2007cf}, we have proposed a stringy realization of the D-term gauge mediation. The minimal setup of the hidden sector (in type IIB string theory) is $N$ D5-branes wrapped around the 2-cycle at the tip of the resolved conifold. The low energy open string spectrum on the D5-branes gives $SU(N) \times U(1)_D$ super Yang-Mills theory, and the size of the 2-cycle serves as a FI term for the $U(1)_D$ gauge symmetry. To realize the SSM(-like) sector, we introduce D7-branes wrapping a holomorphic 4-cycle in the bulk region of the conifold. We assume that the SSM(-like) theory is realized on the D7-branes.
The 5-7 strings can be identified as the messenger in the D-term gauge mediation, and the separation between the D5-branes and D7-branes gives supersymmetric mass for the messenger. The matter content is exactly what we need for the non-perturbative D-term gauge mediation.

In \cite{Nakayama:2007cf}, we assumed the stabilization of the resolution parameter of the conifold  (FI parameter) by focusing on the local geometry. In the full compactification, it will be stabilized by flux and nonperturbative effects. Although the detailed mechanism of the moduli stabilization is beyond the scope of this paper, the important point is that the stabilization of the dynamical FI term requires the F-term SUSY breaking as well.\footnote{See e.g. \cite{Nakayama:2007du}\cite{NP} for some attempts to fix the FI parameter in the metastable SUSY breaking models realized in the string theory.} This F-term could couple with the $U(1)'$ gauge symmetry realized on D7-branes through the R-R exchange mechanism as has been studied in \cite{Verlinde:2007qk}.\footnote{If we realize the SSM sector by using the D5-branes sitting at another tip of the cone inside the total Calabi-Yau space, the discussion in \cite{Verlinde:2007qk} takes over to our case word by word. One (massive) extra $U(1)$ gauge group would be identified with $U(1)_D$ and the light $U(1)$ gauge group would be identified with $U(1)'$. The two-cycles on which the hidden sector D-brane wraps and SSM D-brane wraps should be identical in cohomology inside the total Calabi-Yau space but be locally distinguished.} 

The realization of the D-term SUSY breaking in the string theory has been studied in various context recently \cite{Dymarsky:2005xt}\cite{Aganagic:2007zm}. The explicit construction of the combined D-term gauge mediation and $U(1)'$ mediation in such models would be of great interest. We also note that the extra $U(1)'$ contribution discussed here has wider applications: it could be used as a solution to the possible negative SSM sfermion mass squared of the strongly coupled F-term gauge mediation \cite{Izawa:2005yf}\cite{Ibe:2007wp}\cite{INY}. The stringy construction of such strongly coupled F-term gauge mediation would be interesting as well.

\section*{Acknowledgements}
The author would like to thank M.~Ibe, P.~Kumar, M.~Yamazaki, and T.~T.~Yanagida for valuable discussions.
The research of Y.~N. is supported in part by NSF grant PHY-0555662 and the UC Berkeley Center for Theoretical Physics.

\appendix
\section{Stable $U(1)'$ {\it gauge} mediation}
As discussed in the main text, the origin of the interaction \eqref{mass} is crucial to understand the stability of our model under the thermal effects. In this appendix, we show a possible microscopic origin of the gaugino mass from the stable $U(1)'$ {\it gauge} mediation. A major challenge here is how to avoid the emergence of the SUSY preserving vacua.
We also discuss a small subtlety of the effective SSM sfermion mass squared formula used in the main text.

We assume that the SUSY breaking chiral superfield $S$ in the hidden sector has a shift symmetry $S \to S + 2iM_0 x$ in the imaginary direction (i.e. $x\in \mathbf{R}$). We also introduce $U(1)'$ messenger chiral superfield $\varphi$ and $\bar{\varphi}$ transforming linearly (say $\varphi \to e^{-ix}\varphi$, $\bar{\varphi}\to e^{-ix}\bar{\varphi}$) under the shift symmetry so that $e^{S/M_0} \varphi\bar{\varphi}$ is invariant.\footnote{Because of the $U(1)$ anomaly, the possibility of stable GUT messenger is excluded.} The total superpotential invariant under the shift symmetry (after the superspace integration)\footnote{We could also introduce a bare $S$ dependent gauge kinetic term $c S/M_{0} W^\alpha W_\alpha$ without violating the shift symmetry, but as long as $c$ is small, the effect is negligible.}  is
\begin{eqnarray}
W = \Lambda_S^2S + m_0e^{S/M_0} \varphi \bar{\varphi} \ .
\end{eqnarray}
We set the Kahler potential so that $\langle \mathrm{Rr} S \rangle =0$ is the stable vacuum. 
Importantly, the model does not have any SUSY vacuum. The stability of the messenger requires $F_S < M_0 m_0$, which will be satisfied e.g. by taking $m_0 \sim M_{pl}$.

The usual gauge mediation formula gives the $U(1)'$ gaugino mass as
\begin{eqnarray}
m_{1/2;Z'} \sim \frac{g_{Z'}^2}{16\pi^2} \frac{\Lambda_S^2}{M_0} \ .
\end{eqnarray}
Furthermore, the SSM sfermions obtain soft mass squared at the two-loop order as
\begin{eqnarray}
m_i^2 \sim \left(\frac{g_{Z'}^2}{16\pi^2}  \frac{\Lambda_S^2}{M_0}\right)^2 \ .\label{offset}
\end{eqnarray}
Note that the formula \eqref{offset} is different than \eqref{xsfm} taken from \cite{Langacker:2007ac}. The reason is that in the $U(1)'$ {\it gauge} mediation, in contrast to the  scenario discussed in the main text, after integrating out the messenger, we have an effective K\"ahler interaction $\int d^4\theta \frac{g_{Z'}^4}{(16\pi^2)^2m_0^2}(S+S^\dagger)^2 \Phi_{SM} \Phi_{SM}^\dagger$ in addition to the $U(1)'$ gaugino mass  $\int d^2\theta \frac{S}{m} W^\alpha W_\alpha$ discussed in the main text. The contribution \eqref{offset} dominates over  \eqref{xsfm}, but they are not considered in \cite{Langacker:2007ac} because they did not specify the origin of all these effective non-renormalizable interactions.

The subsequent analysis can be done completely in line with the discussion in section 3. We take $M_0 \sim \Lambda_S$ to realize $O(10)$ eV gravitino mass. The D-term gauge mediation gives rise to the SSM gaugino mass. A possible negative mass squared for the SSM sfermions can be offset by  \eqref{offset}. One phenomenological prediction of this $U(1)'$ {\it gauge} mediation is that the $Z'$ boson is as light as the SUSY particles.


\begin{thebibliography}{99}
%\cite{Pagels:1981ke}
\bibitem{Pagels:1981ke}
  H.~Pagels and J.~R.~Primack,
  %``Supersymmetry, Cosmology And New Tev Physics,''
  Phys.\ Rev.\ Lett.\  {\bf 48}, 223 (1982).
  %%CITATION = PRLTA,48,223;%%
%\cite{Khlopov:1984pf}
\bibitem{K1}
  M.~Y.~Khlopov and A.~D.~Linde,
  %``Is It Easy To Save The Gravitino?,''
  Phys.\ Lett.\  B {\bf 138} (1984) 265.
  %%CITATION = PHLTA,B138,265;%%

%\cite{Falomkin:1984eu}
\bibitem{K2}
  I.~V.~Falomkin, G.~B.~Pontecorvo, M.~G.~Sapozhnikov, M.~Y.~Khlopov, F.~Balestra and G.~Piragino,
  %``Low-Energy Anti-P He-4 Annihilation And Problems Of The Modern Cosmology,
  %GUT And Susy Models,''
  Nuovo Cim.\  A {\bf 79} (1984) 193
  [Yad.\ Fiz.\  {\bf 39} (1984) 990].
  %%CITATION = NUCIA,79A,193;%%

%\cite{Moroi:1993mb}
\bibitem{Moroi:1993mb}
  T.~Moroi, H.~Murayama and M.~Yamaguchi,
  %``Cosmological constraints on the light stable gravitino,''
  Phys.\ Lett.\  B {\bf 303}, 289 (1993).
  %%CITATION = PHLTA,B303,289;%%
%\cite{Khlopov:1993ye}
\bibitem{K3}
  M.~Y.~Khlopov, Yu.~L.~Levitan, E.~V.~Sedelnikov and I.~M.~Sobol,
  %``Nonequilibrium cosmological nucleosynthesis of light elements: Calculations
  %by the Monte Carlo method,''
  Phys.\ Atom.\ Nucl.\  {\bf 57}, 1393 (1994)
  [Yad.\ Fiz.\  {\bf 57}, 1466 (1994)].
  %%CITATION = YAFIA,57,1466;%%

%\cite{de Gouvea:1997tn}
\bibitem{de Gouvea:1997tn}
  A.~de Gouvea, T.~Moroi and H.~Murayama,
  %``Cosmology of supersymmetric models with low-energy gauge mediation,''
  Phys.\ Rev.\  D {\bf 56}, 1281 (1997)
  [arXiv:hep-ph/9701244].
  %%CITATION = PHRVA,D56,1281;%%
%\cite{K4}
\bibitem{K4}
  M.~Y.~Khlopov, A.~Barrau and J.~Grain,
  %``Gravitino production by primordial black hole evaporation and  constraints
  %on the inhomogeneity of the early universe,''
  Class.\ Quant.\ Grav.\  {\bf 23}, 1875 (2006)
  [arXiv:astro-ph/0406621].
  %%CITATION = CQGRD,23,1875;%%
\bibitem{Viel:2005qj}
  M.~Viel, J.~Lesgourgues, M.~G.~Haehnelt, S.~Matarrese and A.~Riotto,
  %``Constraining warm dark matter candidates including sterile neutrinos  and
  %light gravitinos with WMAP and the Lyman-alpha forest,''
  Phys.\ Rev.\  D {\bf 71}, 063534 (2005)
  [arXiv:astro-ph/0501562].
  %%CITATION = PHRVA,D71,063534;%%
%\cite{Izawa:1997gs}
\bibitem{Izawa:1997gs}
  K.~I.~Izawa, Y.~Nomura, K.~Tobe and T.~Yanagida,
  %``Direct-transmission models of dynamical supersymmetry breaking,''
  Phys.\ Rev.\  D {\bf 56}, 2886 (1997)
  [arXiv:hep-ph/9705228].
  %%CITATION = PHRVA,D56,2886;%%
%\cite{Izawa:1999vc}
\bibitem{Izawa:1999vc}
  K.~I.~Izawa, Y.~Nomura and T.~Yanagida,
  %``A gauge-mediation model of dynamical SUSY breaking with a wide range of
  %the gravitino mass,''
  Phys.\ Lett.\  B {\bf 452}, 274 (1999)
  [arXiv:hep-ph/9901345].
  %%CITATION = PHLTA,B452,274;%%
%\cite{Izawa:2005yf}
\bibitem{Izawa:2005yf}
  K.~I.~Izawa and T.~Yanagida,
  %``Strongly coupled gauge mediation,''
  Prog.\ Theor.\ Phys.\  {\bf 114}, 433 (2005)
  [arXiv:hep-ph/0501254].
  %%CITATION = PTPKA,114,433;%%
%\cite{Ibe:2007wp}
\bibitem{Ibe:2007wp}
  M.~Ibe, Y.~Nakayama and T.~T.~Yanagida,
  %``Conformal gauge mediation,''
  Phys.\ Lett.\  B {\bf 649}, 292 (2007)
  [arXiv:hep-ph/0703110].
  %%CITATION = PHLTA,B649,292;%%
%\cite{Viel:2005qj}

%\cite{Ibe:2007ab}
\bibitem{Ibe:2007ab}
  M.~Ibe and R.~Kitano,
  %``Minimal Direct Gauge Mediation,''
  arXiv:0711.0416 [hep-ph].
  %%CITATION = ARXIV:0711.0416;%%
%\cite{INY}
\bibitem{INY}
M.~Ibe, Y.~Nakayama and T.~T.~Yanagida,
to appear.




%\cite{Nakayama:2007cf}
\bibitem{Nakayama:2007cf}
  Y.~Nakayama, M.~Taki, T.~Watari and T.~T.~Yanagida,
  %``Gauge mediation with D-term SUSY breaking,''
  Phys.\ Lett.\  B {\bf 655}, 58 (2007)
  [arXiv:0705.0865 [hep-ph]].
  %%CITATION = PHLTA,B655,58;%%

%\cite{Poppitz:1996xw}
\bibitem{Poppitz:1996xw}
  E.~Poppitz and S.~P.~Trivedi,
  %``Some remarks on gauge-mediated supersymmetry breaking,''
  Phys.\ Lett.\  B {\bf 401}, 38 (1997)
  [arXiv:hep-ph/9703246].
  %%CITATION = PHLTA,B401,38;%%

%\cite{Langacker:2007ac}
\bibitem{Langacker:2007ac}
  P.~G.~Langacker, G.~Paz, L.~T.~Wang and I.~Yavin,
  %``Z'-mediated Supersymmetry Breaking,''
  arXiv:0710.1632 [hep-ph].
  %%CITATION = ARXIV:0710.1632;%%
%\cite{Blumenhagen:2005mu}
\bibitem{Blumenhagen:2005mu}
  R.~Blumenhagen, M.~Cvetic, P.~Langacker and G.~Shiu,
  %``Toward realistic intersecting D-brane models,''
  Ann.\ Rev.\ Nucl.\ Part.\ Sci.\  {\bf 55}, 71 (2005)
  [arXiv:hep-th/0502005].
  %%CITATION = ARNUA,55,71;%%
%\cite{Blumenhagen:2006ci}
\bibitem{Blumenhagen:2006ci}
  R.~Blumenhagen, B.~Kors, D.~Lust and S.~Stieberger,
  %``Four-dimensional String Compactifications with D-Branes, Orientifolds   and
  %Fluxes,''
  Phys.\ Rept.\  {\bf 445}, 1 (2007)
  [arXiv:hep-th/0610327].
  %%CITATION = PRPLC,445,1;%%

%\cite{Verlinde:2007qk}
\bibitem{Verlinde:2007qk}
  H.~Verlinde, L.~T.~Wang, M.~Wijnholt and I.~Yavin,
  %``A Higher Form (of) Mediation,''
  arXiv:0711.3214 [hep-th].
  %%CITATION = ARXIV:0711.3214;%%

%\cite{Choi:2005ge}
\bibitem{Choi:2005ge}
  K.~Choi, A.~Falkowski, H.~P.~Nilles and M.~Olechowski,
  %``Soft supersymmetry breaking in KKLT flux compactification,''
  Nucl.\ Phys.\  B {\bf 718}, 113 (2005)
  [arXiv:hep-th/0503216].
  %%CITATION = NUPHA,B718,113;%%
\bibitem{NYY2}
Y.~Nakayama, M.~Yamazaki and T.~T.~Yanagida,
to appear.
%\cite{Nakayama:2007du}
\bibitem{Nakayama:2007du}
  Y.~Nakayama, M.~Yamazaki and T.~T.~Yanagida,
  %``Moduli Stabilization in Stringy ISS Models,''
  arXiv:0710.0001 [hep-th].
  %%CITATION = ARXIV:0710.0001;%%
\bibitem{NP}
Y.~Nakayama and P.~Kumar
to appear.

%\cite{Giudice:1998bp}
\bibitem{Giudice:1998bp}
  G.~F.~Giudice and R.~Rattazzi,
  %``Theories with gauge-mediated supersymmetry breaking,''
  Phys.\ Rept.\  {\bf 322}, 419 (1999)
  [arXiv:hep-ph/9801271].
  %%CITATION = PRPLC,322,419;%%


%\cite{Hamaguchi:2007rb}
\bibitem{Hamaguchi:2007rb}
  K.~Hamaguchi, S.~Shirai and T.~T.~Yanagida,
  %``Composite Messenger Baryon as a Cold Dark Matter,''
  Phys.\ Lett.\  B {\bf 654}, 110 (2007)
  [arXiv:0707.2463 [hep-ph]].
  %%CITATION = PHLTA,B654,110;%%

%\cite{Everett:2000hb}
\bibitem{Everett:2000hb}
  L.~L.~Everett, P.~Langacker, M.~Plumacher and J.~Wang,
  %``Alternative supersymmetric spectra,''
  Phys.\ Lett.\  B {\bf 477}, 233 (2000)
  [arXiv:hep-ph/0001073].
  %%CITATION = PHLTA,B477,233;%%

%\cite{Dymarsky:2005xt}
\bibitem{Dymarsky:2005xt}
  A.~Dymarsky, I.~R.~Klebanov and N.~Seiberg,
  %``On the moduli space of the cascading SU(M+p) x SU(p) gauge theory,''
  JHEP {\bf 0601}, 155 (2006)
  [arXiv:hep-th/0511254].
  %%CITATION = JHEPA,0601,155;%%

%\cite{Aganagic:2007zm}
\bibitem{Aganagic:2007zm}
  M.~Aganagic and C.~Beem,
  %``Geometric Transitions and D-Term SUSY Breaking,''
  arXiv:0711.0385 [hep-th].
  %%CITATION = ARXIV:0711.0385;%%


\end{thebibliography}
\end{document}